\documentclass[aps,amsmath,amssymb,groupedaddress,twocolumn]{revtex4}
\usepackage{graphicx}

\begin{document}

\title{Physical limits on information processing}

\author{Stephen~D.~H.~Hsu} \email{hsu@duende.uoregon.edu}
\affiliation{Institute of Theoretical Science \\ University of Oregon,
Eugene, OR 97403}

\begin{abstract}
We derive a fundamental upper bound on the rate at which a device can
process information (i.e., the number of logical operations per unit
time), arising from quantum mechanics and general relativity. In
Planck units a device of volume $V$ can execute no more than the cube
root of $V$ operations per unit time. We compare this to the rate of
information processing performed by nature in the evolution of
physical systems, and find a connection to black hole entropy and the
holographic principle.
\end{abstract}


\maketitle

\date{today}

\bigskip

\bigskip

In this note we derive an upper bound on the rate at which a device
can process information. We define this rate as the number of logical
operations per unit time, denoted as the ops rate ${\cal R}$. The
operations in question can be those of either classical or quantum
computers. The basis of our result can be stated very simply:
information processing requires energy, and general relativity limits
the energy density of any object that does not collapse to a black
hole. Replacing {\it information processing} by {\it information} in
the previous sentence leads to holography or black hole entropy
bounds, a connection we will explore further below. For related work
on fundamental physical limits to computation, see \cite{lloyd} and
\cite{lloydng}. We use Planck units throughout, in which the speed of
light, Planck's constant and the Planck mass (equivalently, Newton's
constant) are unity.

Our result is easily deduced using the Margolus--Levitin (ML) theorem
\cite{lm} from quantum mechanics, and the hoop conjecture from general
relativity, originally formulated by Thorne \cite{hoop}.

The Margolus--Levitin theorem states that a quantum system with
average energy $E$ requires at least $\Delta t > \frac{\pi}{2} E^{-1}$
to evolve into an orthogonal (distinguishable) state. It is easy to
provide a heuristic justification of this result. For an energy
eigenstate of energy $E$, $E^{-1}$ is the time required for its phase
to change by order one. In a two state system the energy level
splitting $E$ is at most of order the average energy of the two
levels. Then, the usual energy-time uncertainty principle suggests
that the system cannot be made to undergo a controlled quantum jump on
timescales much less than $E^{-1}$, as this would introduce energy
larger than the splitting into the system.

The hoop conjecture gives a criteria for gravitational collapse. It
states that a system of total energy $M$, if confined to a sphere of
radius $L < \eta M$ ($\eta$ is a coefficient of order one, which we
neglect below), must inevitably evolve into a black hole. The
condition $L < M$ is readily motivated by the Schwarzschild radius
$R_s = 2M$. This conjecture has been confirmed in
astrophysically-motivated numerical simulations, and has been placed
on even stronger footing by recent results on black hole formation
from relativistic particle collisions \cite{bh}. These results show
that, even in the case when all of the energy $M$ is provided by the
kinetic energy of two highly boosted particles, a black hole forms
whenever the particles pass within a distance of order $M$ of each
other. Two particle collisions had seemed the most likely to provide a
counterexample to the conjecture, since the considerable energy of
each particle might have allowed escape from gravitational
collapse. One can think of the hoop conjecture as requiring that the
average energy density of an object of size $L$ be bounded above by
$L^{-2}$ in order not to collapse to a black hole. Thus, large objects
which are not black holes must be less and less dense.

Our main result follows directly. Consider a device of size $L$ and
volume $V \sim L^3$, comprised of $n$ individual components \cite{fn1}
of average energy $E$. Then, the ML theorem gives an upper bound on
the total number of operations per unit time
\begin{equation}
{\cal R} < n E~,
\end{equation}
while the hoop conjecture requires $M \sim n E < L$. Combined, we obtain
\begin{equation}
\label{bound}
{\cal R} < L \sim V^{1/3}~.
\end{equation}

It is interesting to compare this bound to the rate of information
processing performed by nature in the evolution of physical
systems. At first glance, there appears to be a problem since one
typically assumes the number of degrees of freedom in a region is
proportional to $V$ (is extensive). Then, the amount of information
processing necessary to evolve such a system in time grows much faster
than our bound (\ref{bound}) as $V$ increases. Recall that for $n$
degrees of freedom (for simplicity, qubits), the dimension of Hilbert
space $H$ is $N = {\rm dim} \, H = 2^n$ and the entropy is $S = \ln N
\sim n$. In the extensive case, $n \sim S \sim V$.

However, gravity also constrains the maximum
information content (entropy $S$) of a region of space. 't Hooft
\cite{thooft} showed that if one excludes states from the Hilbert
space whose energies are so large that they would have already caused
gravitational collapse, one obtains $S = \ln N < A^{3/4}$, where $N$
is the number of degrees of freedom and $A$ the surface area.  To
deduce this result, 't Hooft replaces the system under study with a
thermal one. The number of states of a system with constant total
energy $M$ is given to high accuracy by the thermal result in the
large volume limit (recall the relation between the microcanonical and
canonical ensembles in statistical mechanics). Given a thermal region
of radius $L$ and temperature $T$, we have $S \sim T^3 L^3$ and $M
\sim T^4 L^3$. Requiring $M < L$ then implies $T \sim L^{-1/2}$ and $S
< L^{3/2} \sim A^{3/4}$. We stress that the thermal replacement is
just a calculational trick: temperature plays no role in the results,
which can also be obtained by direct counting. In \cite{buniyhsu}, it
was shown that imposing the condition ${\rm Tr} [ \, \rho H \,] < L$
on a density matrix $\rho$ implies a similar bound $S_{\rm vN} <
A^{3/4}$ on the von Neumann entropy $S_{\rm vN} = - {\rm Tr} \, \rho \ln
\rho$. For $\rho$ a pure state the result reduces to the previous
Hilbert space counting.  We note that these bounds are more
restrictive than the bound obtained from black hole entropy: $S < A$
\cite{area}. One can interpret this discrepancy as a consequence of
higher entropy density of gravitational degrees of freedom relative to
ordinary matter \cite{hsumurray}.

Using these results we can calculate the maximum rate of information
processing necessary to simulate any physical system of volume $V$
which is not a black hole. The rate ${\cal R}$ is given by the number
of degrees of freedom $S \sim L^{3/2}$ times the maximal ML rate $T
\sim L^{-1/2}$. This yields ${\cal R} \sim L$ as in our bound
(\ref{bound}).

Finally, we note that black holes themselves appear to saturate our
bound. If we take the black hole entropy to be $S \sim A \sim L^2$,
and the typical energy of its modes to be the Hawking temperature
$T_{\rm H} \sim L^{-1}$, we again obtain ${\cal R} \sim L$.

\bigskip
\emph{Acknowledgements.---} The author thanks R. Buniy, X. Calmet and
A. Zee for comments, and P. Davies for stimulating his interest in
this problem. This work was supported in part under DOE contract
DE-FG06-85ER40224.


\bigskip



\begin{thebibliography}{99}
\bibitem{fn1} We identify $n$ individual components for later
comparison with entropy. The ML theorem could, of course, be applied
to the entire device as well.
\end{thebibliography}
\end{document}